\documentclass[12pt]{article}

\usepackage[left=1in,right=1in,top=1in,bottom=1in]{geometry}
\usepackage{graphicx}
\usepackage{float}
\usepackage{amsfonts}
\usepackage{amssymb}
\usepackage{amsmath}
\usepackage{relsize}
\usepackage{xcolor}

\def\gtwid{\mathrel{\raise.3ex\hbox{$>$\kern-.75em\lower1ex\hbox{$\sim$}}}}
\def\ltwid{\mathrel{\raise.3ex\hbox{$<$\kern-.75em\lower1ex\hbox{$\sim$}}}}
\def\square{\kern1pt\vbox{\hrule height 1.2pt\hbox{\vrule width 1.2pt\hskip 3pt
   \vbox{\vskip 6pt}\hskip 3pt\vrule width 0.6pt}\hrule height 0.6pt}\kern1pt}

\begin{document}

\begin{titlepage}

\begin{flushright}
UFIFT-QG-23-03
\end{flushright}

\vskip 4cm

\begin{center}
{\bf Unfinished Business in A Nonlinear Sigma Model on de Sitter Background}
\end{center}

\vskip 1cm

\begin{center}
R. P. Woodard$^{\dagger}$ and B. Yesilyurt$^{*}$
\end{center}

\begin{center}
\it{Department of Physics, University of Florida, \\
Gainesville, FL 32611, UNITED STATES}
\end{center}

\vspace{1cm}

\begin{center}
ABSTRACT
\end{center}
Nonlinear sigma models on de Sitter background possess the same kind of
derivative interactions as gravity, and show the same sorts of large 
spacetime logarithms in correlation functions and solutions to the 
effective field equations. It was recently demonstrated that these logarithms
can be resummed by combining a variant of Starobinsky's stochastic 
formalism with a variant of the renormalization group. This work 
considers one of these models and completes two pieces of analysis 
which were left unfinished: the evolution of the background at two 
loop order and the one loop beta function. 

\begin{flushleft}
\end{flushleft}

\vspace{3cm}

\begin{flushleft}
$^{\dagger}$ e-mail: woodard@phys.ufl.edu \\
$^{*}$ e-mail: b.yesilyurt@ufl.edu
\end{flushleft}

\end{titlepage}

\section{Introduction}

In addition to proposing one of the first theories of primordial inflation
\cite{Starobinsky:1980te}, Starobinsky was the first to realize that the
accelerated expansion of inflation generates a spectrum of gravitational 
radiation \cite{Starobinsky:1979ty}. The numbers are staggering. If $a(\eta)$ 
is the scale factor at conformal time $\eta$, and $\Delta^2_{h}(k)$ is the 
tensor power spectrum, then the number of inflationary gravitons with 
co-moving wave number $k$ and each $\pm$ polarization is,
\begin{equation}
N_{\pm}(\eta,k) = \frac{\pi \Delta^2_{h}(k)}{64 G k^2} \!\times\! a^2(\eta)
\longrightarrow \Bigl( \frac{H a}{2 k} \Bigr)^2 \; . \label{Ngrav}
\end{equation}
Notice that the left hand form is valid for a general inflationary background 
whereas the right hand form is specialized to the de Sitter geometry,
\begin{equation}
ds^2 \equiv g_{\mu\nu} dx^{\mu} dx^{\nu} = a^2 \Bigl[ -d\eta^2 + d\vec{x} 
\!\cdot\! d\vec{x}\Bigr] \qquad , \qquad a(\eta) = -\frac1{ H \eta} \; . 
\label{deSitter}
\end{equation}
In the distant past, when $a \simeq 0$, the occupation number (\ref{Ngrav}) is 
nearly zero. At horizon crossing it becomes $\frac14$, and within a single 
e-foldings it has exceeded unity. Solving the horizon problem requires at least 
50 e-foldings of inflation, which causes the occupation number to expand by a 
factor of $e^{100} \simeq 2.6 \times 10^{43}$. One should bear in mind that 
this is just the occupation number of a single wave vector $\vec{k}$. Of course
the 3-volume is expanding like $a^3(t)$. Integrating over all super-horizon 
modes gives a constant number density of $\frac{H^3}{8 \pi^2}$, or $0.01$ 
distinct super-horizon modes for each Hubble volume. 

Loops of these inflationary gravitons change the kinematics and forces
exerted by themselves and by other particles. One shows this by 
quantum correcting the effective field equation of whatever particle 
is under study using the graviton contribution to its 1PI 
(one-particle-irreducible) 2-point function. For example, the 1PI 
2-point function of photons is a bi-vector density known as the 
vacuum polarization $i[\mbox{}^{\mu} \Pi^{\nu}](x;x')$, and was first
computed, at 1-graviton loop order, on de Sitter background in 2013 
\cite{Leonard:2013xsa} using the simplest graviton gauge 
\cite{Tsamis:1992xa,Woodard:2004ut}. The quantum-corrected Maxwell 
equation is,
\begin{equation}
\partial_{\nu} \Bigl[ \sqrt{-g} \, g^{\nu\rho} g^{\mu\sigma} 
F_{\rho\sigma} \Bigr] + \int \!\! d^4x' \Bigl[\mbox{}^{\mu} \Pi^{\nu}
\Bigr](x;x') A_{\nu}(x') = J^{\mu}(x) \; , \label{QMax}
\end{equation}
where $F_{\mu\nu} \equiv \partial_{\mu} A_{\nu} - \partial_{\nu} 
A_{\mu}$ is the field strength tensor and $J^{\mu}$ is the current
density. When (\ref{QMax}) is solved for a static point charge $Q$, 
the resulting Coulomb potential is \cite{Glavan:2013jca},
\begin{equation}
\Phi(\eta,r) = \frac{Q}{4\pi a r} \Biggl\{ 1 + \frac{2 G}{3 \pi a^2 r^2}
+ \frac{2 G H^2}{\pi} \times \ln(a H r) + O(G^2) \Biggr\} . 
\label{Coulomb}
\end{equation}
Solving (\ref{QMax}) for a dynamical photon shows a similar 1-loop
enhancement of the tree order electric field strength $F^{0i}_{0}$
\cite{Wang:2014tza},
\begin{equation}
F^{0i}(\eta,\vec{x}) = F^{0i}_{0}(\eta,\vec{x}) \Biggl\{1 + 
\frac{2 G H^2}{\pi} \times \ln(a) + O(G^2) \Biggr\} . \label{FStrength}
\end{equation}

Many similar calculations have been made in recent years. Single graviton 
loop contributions on de Sitter background have been evaluated for the 1PI 
2-point functions of gravitons \cite{Tsamis:1996qk}, fermions 
\cite{Miao:2005am,Miao:2012bj}, massless, minimally coupled scalars 
\cite{Kahya:2007bc} and massless, conformally coupled scalars 
\cite{Boran:2014xpa,Boran:2017fsx,Glavan:2020gal}. Large temporal and 
spatial logarithms analogous to (\ref{Coulomb}-\ref{FStrength}) have been 
found for the fermion field strength \cite{Miao:2006gj}, the massless, 
minimally coupled scalar exchange potential \cite{Glavan:2021adm}, the 
graviton mode function \cite{Tan:2021lza}, and the Newtonian potential of 
a static point mass \cite{Tan:2022xpn}.

A fascinating feature of these results is the breakdown of perturbation
theory that occurs when the large logarithms overwhelm the small loop-counting
parameter of $G H^2 \ltwid 10^{-10}$. Graviton loop corrections do not necessarily 
become large at this point because each higher loop contributes an extra factor
of the loop-counting parameter times a large temporal or spatial logarithm, and
they all become of order one at the same time. Understanding what happens after 
the breakdown of perturbation theory requires a way of resumming the series of
leading logarithms. Although Starobinsky's stochastic formalism 
\cite{Starobinsky:1986fx,Starobinsky:1994bd} solves the analogous problem for 
the large logarithms of scalar potential models \cite{Woodard:2005cw,
Tsamis:2005hd}, the derivative interactions of quantum gravity cause it to fail
\cite{Miao:2008sp}. The renormalization group is another obvious approach but
the simplest realization of it fails even to describe scalar potential models
\cite{Woodard:2008yt}. This has remained true despite years of effort 
\cite{Burgess:2009bs,Burgess:2010dd,Burgess:2015ajz}.

Nonlinear sigma models have the same derivative interactions as quantum gravity
but without the indices and the gauge issue. It has long been suspected that 
they might provide a simplified venue for working out a procedure to sum the 
leading logarithms \cite{Tsamis:2005hd}, and much work has been done with them
on de Sitter background \cite{Kitamoto:2010et,Kitamoto:2011yx,Kitamoto:2018dek}.
These efforts have recently resulted in a synthesis involving variants of 
Starobinsky's stochastic technique with the renormalization group 
\cite{Miao:2021gic}. What one does is to derive curvature-dependent effective
potentials by integrating derivative interactions out of the field equations in
the presence of a constant scalar background. The resulting equation is that of
a scalar potential model for which Starobinsky's stochastic procedure applies
\cite{Starobinsky:1986fx,Starobinsky:1994bd}. This recovers some large logarithms;
the remaining ones come by employing the renormalization group on the subset of 
BPHZ (Bogoliubov-Parasiuk-Hepp-Zimmermann) counterterms which can be regarded as 
curvature-dependent renormalizations of parameters in the original theory.

The purpose of this paper is to complete two parts of the previous analysis 
\cite{Miao:2021gic}. First, we compute the evolution of the scalar background 
at two loop order. This permits a nontrivial check of the stochastic prediction,
which really is a prediction because the explicit calculation had not been done
when it was made. The second thing we do here is to compute the 1-loop beta 
function, which was not needed to check the 1-loop exchange potentials. Having
the beta function allows us to carry out a full RG analysis.

There are five sections in this paper. In section 2 we define the particular
nonlinear sigma model to be studied and give the stochastic prediction for 
its background. The 2-loop expectation value of the background is computed 
in section 3. In section 4 we work out the 1-loop beta function. Our conclusions
comprise section 5.

\section{Feynman Rules}

The purpose of this section is to explain how to make perturbative computations in
the model. We begin by giving the bare Lagrangian, the propagators and the vertices.
Then we present those counterterms which are needed for our work. The special
counterterms which can be viewed as curvature-dependent renormalizations of the
bare theory are distinguished. We next using a curvature-dependent effective 
potential to derive the stochastic prediction for the 2-loop expectation value 
which is computed in section 3. The section closes with a review of what has been
shown in previous work, and what will be shown in this paper.

\subsection{The Bare Theory}

The Lagrangian we study consists of two scalar fields $A(x)$ and $B(x)$. Their
Lagrangian is,
\begin{equation}
\mathcal{L} = -\frac12 \partial_{\mu} A \partial_{\nu} A g^{\mu\nu} \sqrt{-g} -
\frac12 \Bigl(1 \!+\! \frac12 \lambda A\Bigr)^2 \partial_{\mu} B \partial_{\nu} B
g^{\mu\nu} \sqrt{-g} \; . \label{Lagrangian}
\end{equation}
We work on $D$-dimensional spacetime in order to facilitate the use of dimensional
regularization. Our notation exploits the conformal coordinate system evident in
the background geometry (\ref{deSitter}). Because the metric consists of a scale
factor $a(\eta) = -\frac1{H \eta}$ times the Minkowski metric, $g_{\mu\nu} = a^2
\eta_{\mu\nu}$, its inverse is $g^{\mu\nu} = \frac1{a^2} \eta^{\mu\nu}$, and the 
measure factor is $\sqrt{-g} = a^D$. The standard partial derivative is denoted
as $\partial_{\mu}$, no matter what tensor it acts upon, and its index is raised 
and lowered with the Minkowski metric, $\partial^{\mu} \equiv \eta^{\mu\nu}
\partial_{\nu}$. The Minkowski contraction of the partial derivative with itself 
is written $\partial^2 \equiv \eta^{\mu\nu} \partial_{\mu} \partial_{\nu}$, again 
no matter what tensor it acts upon. When more than one coordinate is present we 
indicate which one is differentiated by adding a superscript or a subscript, as 
in $\partial^y_{\mu} \equiv \frac{\partial}{\partial y^{\mu}}$ and $\partial^{\mu}_{z} 
\equiv \eta^{\mu\nu} \frac{\partial}{\partial z^{\nu}}$. The same scheme is sometimes 
applied to the metric, as in $g^{\mu\nu}_{y} \equiv g^{\mu\nu}(y)$ and $\sqrt{-g_z} 
\equiv \sqrt{-g(z)}$. It can be employed as well for the scale factor, as in $a_y 
\equiv a(y^0)$ and $a_z \equiv a(z^0)$.

The propagators of both fields obey, 
\begin{equation}
\partial_{\mu} \Bigl[ \sqrt{-g} \, {g}^{\mu\nu} \partial_{\nu} i\Delta(x;x')\Bigr] 
= \partial^{\mu} \Bigl[a^{D-2} \partial_{\mu} i\Delta(x;x')\Bigr] \equiv
\mathcal{D} i\Delta(x;x') = i\delta^D(x \!-\! x') \; . \label{propeqn}
\end{equation}
The homogeneous and isotropic solution is \cite{Onemli:2002hr,Onemli:2004mb},
\begin{equation}
i\Delta(x;x') = F\Bigl( \mathcal{Y}(x;x')\Bigr) + k \ln \Bigl(a(\eta) a(\eta') 
\Bigr) \qquad , \qquad k \equiv \frac{H^{D-2}}{(4\pi)^{\frac{D}2}} 
\frac{\Gamma(D \!-\! 1)}{\Gamma(\frac{D}2)} \; . \label{propagator}
\end{equation}
Here $\mathcal{Y}(x;x') \equiv a(\eta) a(\eta') H^2 \Delta x^2(x;x') \equiv a(\eta)
a(\eta') H^2 (x - x')^{\mu} (x - x')^{\nu} \eta_{\mu\nu}$ is the de Sitter length 
function and the first derivative of the function $F(\mathcal{Y})$ is,
\begin{eqnarray}
\lefteqn{F'(\mathcal{Y}) = -\frac{H^{D-2}}{4 (4 \pi)^{\frac{D}2}} \Biggl\{ \Gamma\Bigl(
\frac{D}2\Bigr) \Bigl( \frac{4}{\mathcal{Y}}\Bigr)^{\frac{D}2} + \Gamma\Bigl( \frac{D}2 
\!+\! 1\Bigr) \Bigl( \frac{4}{\mathcal{Y}}\Bigr)^{\frac{D}2 - 1} } \nonumber \\
& & \hspace{5cm} + \sum_{n=0}^{\infty} \Biggl[ \frac{\Gamma(n \!+\! 
\frac{D}2 \!+\! 2)}{\Gamma(n \!+\! 3)} \Bigl( \frac{\mathcal{Y}}{4}\Bigr)^{n-\frac{D}2 + 2} 
- \frac{\Gamma(n \!+\! D)}{\Gamma(n \!+\! \frac{D}2 \!+\! 1)} \Bigl( 
\frac{\mathcal{Y}}{4}\Bigr)^{n} \Biggr] \Biggr\} . \qquad \label{Fprime}
\end{eqnarray}
Using dimensional regulariztion, the coincidence limits of the propagator and its 
first two derivatives are seen to be,
\begin{eqnarray}
i\Delta(x;x) = k \Bigl[-\pi {\rm cot}\Bigl( \frac{D \pi}{2}\Bigr) + 2 \ln(a)\Bigr] 
\qquad & , & \qquad \partial_{\mu} i\Delta(x;x')\Bigl\vert_{x' = x} = k H a 
\delta^0_{~\mu} \; , \qquad \label{propID1} \\
\partial_{\mu} \partial'_{\nu} i\Delta(x;x')\Bigl\vert_{x' = x} = 
-\Bigl(\frac{D\!-\!1}{D} \Bigr) k H^2 g_{\mu\nu} \qquad & , & \qquad \partial_{\mu} 
i\Delta(x;x) = 2 k H a \delta^{0}_{~\mu} \; . \qquad \label{propID2}
\end{eqnarray}

\begin{figure}[H]
\centering
\includegraphics[width=11cm]{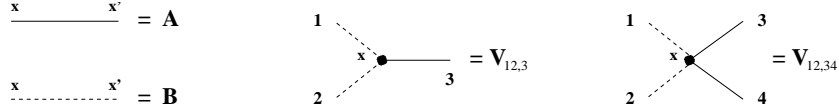}
\caption{\footnotesize The bare Feynman rules of (\ref{Lagrangian}). $A$ lines 
are solid and $B$ lines are dashed. Both propagators are the same and are described
by equations (\ref{propagator}-\ref{Fprime}).}
\label{figAB-1}
\end{figure}

The bare vertices consist of the $\lambda A \partial B \partial B$ coupling and 
the $\lambda^2 A^2 \partial B \partial B$ coupling. Figure~\ref{figAB-1} gives a
diagrammatic representation of the bare Feynman rules.  

\subsection{Counterterms}

\begin{figure}[H]
\centering
\includegraphics[width=5cm]{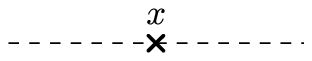}
\hskip 5cm
\includegraphics[width=2.5cm,height=2.5cm]{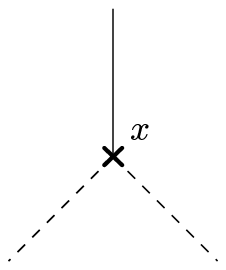}
\caption{\footnotesize Diagrammatic representation of the two counterterms we 
require for this project. The left hand graph represents expression (\ref{B2cterms})
which renormalizes the $B$ self-mass and the right hand graph represents 
(\ref{AB2cterms}) which renormalizes the 3-point vertex. $A$ lines are solid and $B$
lines are dashed.}
\label{Counterterms}
\end{figure}

The Lagrangian (\ref{Lagrangian}) is not renormalizable so it requires an infinite
number of counterterms. However, this paper only employs the two 1-loop counterterms 
which are shown in Figure~\ref{Counterterms}.

The first of the counterterms we require renormalizes the $B$ self-mass at 1-loop 
order,
\begin{equation}
\Delta \mathcal{L}_{B^2} = -\frac12 C_{1B^2} \square B \square B \sqrt{-g} - 
\frac12 C_{2B^2} R \, \partial_{\mu} B \partial_{\nu} B g^{\mu\nu} \sqrt{-g} \; .
\label{B2cterms}
\end{equation}
Absorbing divergences in the $B$ self-mass determines the coefficients 
$C_{1 B^2}$ and $C_{2 B^2}$ to be \cite{Miao:2021gic},
\begin{eqnarray}
C_{1 B^2} & = & -\frac{\lambda^2 \mu^{D-4}}{16 \pi^{\frac{D}2}}
\frac{\Gamma(\frac{D}2 \!-\! 1)}{2 (D\!-\!3) (D\!-\!4)} \; , \label{C1B2} \\
C_{2 B^2} & = & \frac{\lambda^2 \mu^{D-4}}{4 (4\pi)^{\frac{D}2}} 
\frac{\Gamma(D\!-\!1)}{\Gamma(\frac{D}2)} \frac{\pi {\rm cot}(\frac{D \pi}{2})}{
D (D \!-\! 1)} - \frac{\lambda^2 \mu^{D-4}}{32 \pi^{\frac{D}2}} 
\frac{\Gamma(\frac{D}2 \!-\! 1)}{2 (D\!-\!3) (D\!-\!4)} 
\Bigl(\frac{D\!-\!2}{D\!-\!1}\Bigr) \; . \qquad \label{C2B2}
\end{eqnarray}
The term proportional to $C_{1 B^2}$ is a higher derivative counterterm which
plays no role in describing large inflationary logarithms. In contrast, the 
term proportional to $C_{2 B^2}$ can be viewed as a curvature-dependent field 
strength renormalization whose gamma function is,
\begin{equation}
Z_B = 1 + C_{2 B^2} \!\times\! R + O(\lambda^4) \qquad \Longrightarrow \qquad
\gamma_B \equiv \frac{\partial \ln(Z_B)}{\partial \ln(\mu^2)} = 
-\frac{\lambda^2 H^2}{32 \pi^2} + O(\lambda^4) \; . \label{gammaB}
\end{equation} 

The second counterterm renormalizes the 3-point vertex at 1-loop order,
\begin{eqnarray}
\lefteqn{\Delta \mathcal{L}_{AB^2} = -\frac12 C_{1 AB^2} \square A \partial_{\mu} B
\partial_{\nu} B g^{\mu\nu} \sqrt{-g} - C_{2 AB^2} \partial_{\mu} A \partial_{\nu} B
\square B g^{\mu\nu} \sqrt{-g} } \nonumber \\
& & \hspace{4cm} -\frac12 C_{3 AB^2} A \square B \square B \sqrt{-g} - \frac12 
C_{4 AB^2} R \, A \partial_{\mu} B \partial_{\nu} B g^{\mu\nu} \sqrt{-g} \; . 
\qquad \label{AB2cterms}
\end{eqnarray}
The terms involving $C_{1 AB^2}$, $C_{2 AB^2}$ and $C_{3 AB^2}$ contain higher
derivatives and have nothing to do with large inflationary logarithms. The term 
proportional to $C_{4 AB^2}$ can be viewed as a curvature-dependent 
renormalization of the $A \partial B \partial B$ vertex whose beta function 
we will compute at order $\lambda^3$ in section 5.
\begin{equation}
\delta \lambda = C_{4 AB^2} \!\times\! R + O(\lambda^5) \qquad \Longrightarrow
\qquad \beta \equiv \frac{\partial \delta \lambda}{\partial \ln(\mu)} \; .
\label{beta}
\end{equation}

\subsection{Stochastic Prediction for $\langle A \rangle$}

The Heisenberg operator equation for $A(x)$ is,
\begin{equation}
\frac{\delta S}{\delta A} = \partial_{\mu} \Bigl[ \sqrt{-g} g^{\mu\nu}
\partial_{\nu} A\Bigr] - \frac12 \lambda \Bigl(1 \!+\! \frac12 \lambda A\Bigr)
\partial_{\mu} B \partial_{\nu} B g^{\mu\nu} \sqrt{-g} = 0 \; . \label{HEOM}
\end{equation}
A constant $A$ field merely changes the field strength of $B$. We can 
therefore capture the effect of undifferentiated $A$ fields to all orders by
integrating out the differentiated $B$ fields in equation (\ref{HEOM}) for
constant $A$,
\begin{equation}
\partial_{\mu} B \partial_{\nu} B \longrightarrow \frac{ \partial_{\mu} 
\partial'_{\nu} i\Delta(x;x') \vert_{x' \rightarrow x}}{(1 + \frac{\lambda}{2} A)^2}
= -\frac{ (\frac{D-1}{D}) k H^2 g_{\mu\nu} }{(1 + \frac{\lambda}{2} A)^2} 
\longrightarrow -\frac{\frac{3 H^4}{32 \pi^2} g_{\mu\nu}}{(1 + \frac{\lambda}{2}
A)^2} \; . \label{intoutB}
\end{equation}
Note that we have used expression (\ref{propID2}) and the $D=4$ limit of $k$ 
from (\ref{propagator}). 

Substituting (\ref{intoutB}) in the $A$ field equation (\ref{HEOM}) gives a
scalar potential model which Starobinsky's formalism \cite{Starobinsky:1986fx,
Starobinsky:1994bd} allows us to describe, at leading logarithm order, by a
stochastic field $\mathcal{A}$ obeying the Langevin equation \cite{Miao:2021gic},
\begin{equation}
\dot{\mathcal{A}} - \dot{\mathcal{A}}_0 = \frac{ \frac{\lambda H^3}{16 \pi^2}}{
1 + \frac{\lambda}{2} \mathcal{A}} \; . \label{Langevin}
\end{equation}
Note that we have converted from conformal time $\eta$ to the co-moving time
$t = \ln(a)/H$, and dot means derivative with respect to co-moving time. 
The stochastic jitter $\dot{\mathcal{A}}_0$ is provided by the time derivative 
of the infrared truncated, free field mode sum,
\begin{equation}
\mathcal{A}_0(t,\vec{x}) \equiv \int \!\! \frac{d^3k}{(2\pi)^3} \frac{\theta(k \!-\! H)
\theta(a H \!-\! k) H}{\sqrt{2 k^3}} \, \Bigl\{ \alpha_{\vec{k}} \, e^{i \vec{k} 
\cdot \vec{x}} + \alpha^{\dagger}_{\vec{k}} \, e^{-i \vec{k} \cdot \vec{x}} \Bigr\} .
\label{truncfree}
\end{equation}
Here $\alpha^{\dagger}_{\vec{k}}$ and $\alpha_{\vec{k}}$ are the creation and
annihilation operators of the field $A$,
\begin{equation}
\Bigl[ \alpha_{\vec{k}} , \alpha^{\dagger}_{\vec{p}} \Bigr] = (2\pi)^3 \delta^3(
\vec{k} \!-\! \vec{p}) \quad , \quad \alpha_{\vec{k}} \Bigl\vert \Omega 
\Bigr\rangle = 0 \qquad \Longrightarrow \qquad \Bigl\langle \Omega \Bigl\vert 
\mathcal{A}^2_0 \Bigr\vert \Omega \Bigr\rangle = \frac{H^2 \ln(a)}{4 \pi^2} \; . 
\label{creation}
\end{equation} 

Even if the stochastic jitter were absent, $\mathcal{A}(t,\vec{x})$ would roll
down the curvature-dependent effective potential,
\begin{equation}
V_{\rm eff}(A) = -\frac{3 H^4}{8 \pi^2} \, \ln\Bigl\vert 1 + \frac{\lambda}{2} A
\Bigr\vert \; . \label{Veff}
\end{equation}
If one starts from rest at $A = 0$ the exact solution for this behavior is,
\begin{equation}
A_{\rm class} = \frac{2}{\lambda} \Biggl[ \sqrt{1 \!+\! \frac{\lambda^2 H^2 
\ln(a)}{16 \pi^2} } - 1 \Biggr] \; . \label{Aclass}
\end{equation}
The solution of the Langevin equation (\ref{Langevin}) consists of (\ref{Aclass}) 
plus a series in powers of $\mathcal{A}_0$,
\begin{equation}
\mathcal{A} = A_{\rm class} + \mathcal{A}_0 - \frac{\lambda^2 H^3}{32 \pi^2} 
\int_{0}^{t} \!\! dt' \mathcal{A}_0 + \frac{\lambda^3 H^3}{64 \pi^2} 
\int_{0}^{t} \!\! dt' \mathcal{A}^2_0 + O(\lambda^4) \; . \label{Astoch}
\end{equation}
Taking the expectation value gives,
\begin{equation}
\Bigl\langle \Omega \Bigl\vert \mathcal{A} \Bigr\vert \Omega \Bigr\rangle = 
\frac{\lambda H^2 \ln(a)}{2^4 \pi^2} + \frac{\lambda^3 H^4 \ln^2(a)}{2^{10}
\pi^4} + O(\lambda^5) \; . \label{stochastic}
\end{equation} 
In the next section we will show that the 2-loop expectation value of the full 
field $A(x)$ agrees with this stochastic prediction at leading logarithm order.

\subsection{What Has Been and Will Be Shown}

The title of this paper mentions ``unfinished business'' because it completes
an initial investigation of the model (\ref{Lagrangian}) \cite{Miao:2021gic}.
The earlier study computed the 1-loop self-mass for each field and then used
them to quantum-correct the effective field equations to solve for 1-loop
corrections to the plane wave mode functions and to the response to a static
point source. The fields and their squares were also computed at 2-loop order,
with the exception of $A(x)$, whose expectation value was only worked out to
1-loop order. The leading logarithm contributions to the various results are 
summarized in Table~\ref{LargeLogs} below.
\begin{table}[H]
\setlength{\tabcolsep}{8pt}
\def\arraystretch{1.5}
\centering
\begin{tabular}{|@{\hskip 1mm }c@{\hskip 1mm }||c|}
\hline
Quantity & Leading Logarithms \\
\hline\hline
$u_{A}(\eta,k)$ & $\Bigl\{1 {\color{red} -\frac{\lambda^2 H^2}{32 \pi^2} 
\ln(a) } + O(\lambda^4)\Bigr\} \times \frac{H}{\sqrt{2 k^3}}$ \\
\hline
$u_{B}(\eta,k)$ & $\Bigl\{1 + 0 + O(\lambda^4)\Bigr\} \times 
\frac{H}{\sqrt{2 k^3}}$ \\
\hline
$P_{A}(\eta,r)$ & $\Bigl\{1 {\color{red} - \frac{\lambda^2 H^2}{32 \pi^2} 
\ln(a) } {\color{green} + \frac{\lambda^2 H^2}{32 \pi^2} \ln(Hr)} + 
O(\lambda^4)\Bigr\} \times \frac{KH}{4\pi} \ln(Hr)$ \\
\hline
$P_{B}(\eta,r)$ & $\Bigl\{1 {\color{green} - \frac{\lambda^2 H^2}{32 \pi^2} 
\ln(Hr) } + O(\lambda^4)\Bigr\} \times \frac{KH}{4\pi} \ln(Hr)$ \\
\hline
$\langle \Omega \vert A(x)\vert \Omega \rangle$ & $\Bigl\{1 + 
O(\lambda^2)\Bigr\} \times {\color{red} \frac{\lambda H^2}{16 \pi^2} 
\ln(a)}$ \\
\hline
$\langle \Omega \vert A^2(x)\vert \Omega \rangle_{\rm ren}$ & 
$\Bigl\{1 {\color{red} - \frac{\lambda^2 H^2}{64 \pi^2} \ln(a) }
+ O(\lambda^4)\Bigr\} \times {\color{red} \frac{H^2}{4\pi^2} \ln(a)}$ \\
\hline
$\langle \Omega \vert B(x)\vert \Omega \rangle$ & $0$ \\
\hline
$\langle \Omega \vert B^2(x)\vert \Omega \rangle_{\rm ren}$ & 
$\Bigl\{1 {\color{green} + \frac{3 \lambda^2 H^2}{32 \pi^2} \ln(a)}
+ O(\lambda^4) \Bigr\} \times {\color{red} \frac{H^2}{4\pi^2} \ln(a)}$ \\
\hline
\end{tabular}
\caption{\footnotesize Color-coded explanations of large logarithms
found in studying the 1-loop mode functions $u_{A,B}(\eta,k)$, the
1-loop exchange potentials $P_{A,B}(\eta,r)$ and the 2-loop expectation
values of the fields and their squares \cite{Miao:2021gic}. Red denotes 
logarithms which have a stochastic explanation and green indicates those 
explained by the renormalization group.}
\label{LargeLogs}
\end{table}

As Table~\ref{LargeLogs} indicates, the various leading logarithms can
all be explained using a variant of Starobinsky's stochastic formalism ---
based on curvature-dependent effective potentials as explained in section 
2.3 --- or using a variant of the renormalization group --- based on
curvature-dependent renormalizations of the bare theory, as explained in
section 2.2. The goal of this paper is to derive the 2-loop (order $\lambda^3$)
contribution to the expectation value of $A(x)$ --- which will be done in 
section 3 --- and to complete the renormalization group analysis by deriving
the 1-loop beta function --- which will be done in section 4. Under the usual
assumption of no significant higher loop contributions, we will be able to
solve the Callan-Symanzik equation.

\section{The 2-Loop Expectation Value of $A(x)$}

\begin{figure}[H]
\centering
\includegraphics[width=16cm]{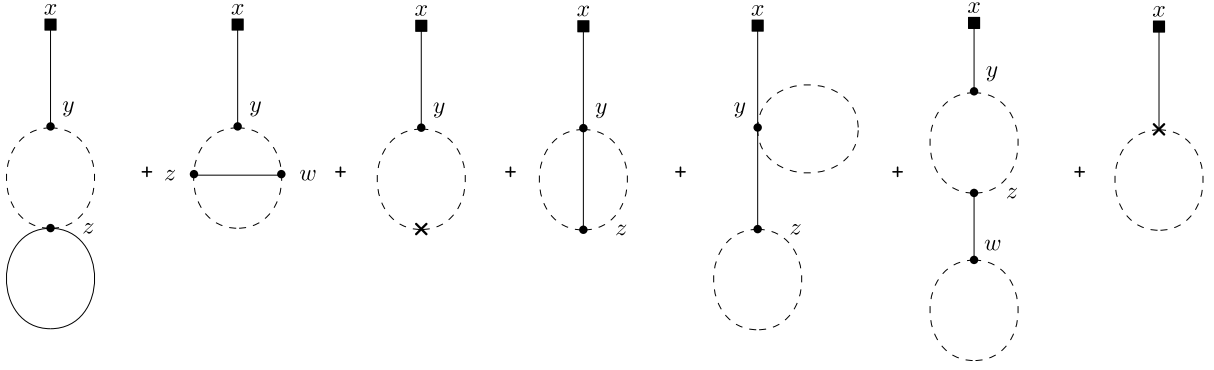}
\caption{\footnotesize Diagrams contributing to the 2-loop expectation value
of $A(x)$. Solid lines represent the $A$ propagator whereas dashed lines 
stand for the $B$ propagators.}
\label{alldiag}
\end{figure}

The point of this section is to compute the expectation value of $A(x)$ at 
2-loop order, and to compare the leading logarithm result for this with the 
stochastic prediction (\ref{stochastic}). The various diagrams are given in 
Figure~\ref{alldiag}.
\noindent For each diagram we give the initial expression. An explicit 
partial integration reduction is presented for the first diagram, and the 
final reductions for each diagram are expressed in terms of $i\Delta(x;x)$,
$[i\Delta(x;x)]^2$ and five integrals,
\begin{eqnarray}
& I_1 & \equiv  \int \!\! d^Dy \sqrt{-g(y)}  \Bigl[i\Delta(x;y) \Bigr]^2 
\; , \qquad \label{I1def} \\
&I_2& \equiv \int \!\! d^Dy \sqrt{-g(y)} \, i\Delta(x;y) i \Delta(y;y) 
\ , \ I_3 \equiv \int \!\! d^Dy \sqrt{-g(y)} \, i\Delta(x;y) \; , \qquad 
\label{I23def} \\
&I_4& \equiv \int \!\! d^Dy \sqrt{-g(y)} \, i\Delta(x;y) \!\times\! I_3(y) 
 \ , \ I_5 \equiv \int \!\! d^Dy \sqrt{-g(y)} \, i\Delta(x;y) \!\times\!
\dot{I}_3(y) \; . \qquad \label{I45def}
\end{eqnarray}
(Note the co-moving time derivative $\partial_t \equiv 1/a \times \partial_0$ 
of $I_3$ in $I_5$.) The proper treatment of these integrals requires generalizing 
in-out matrix elements to true expectation values using the Schwinger-Keldysh 
formalism \cite{Schwinger:1960qe,Mahanthappa:1962ex,Bakshi:1962dv,Bakshi:1963bn,
Keldysh:1964ud} and is explained in the Appendix. The section closes by 
expressing the leading logarithm contribution from each diagram as a factor 
times the 2-loop stochastic result $S \equiv \lambda^3 H^4 \ln^2(a)/2^{10} 
\pi^4$.

\subsection{Computation of the Diagrams}
The first 2-loop diagram in Figure~\ref{alldiag} has a symmetry 
factor of $\frac14$, and a combination of 3-point and 4-point vertices,
\begin{eqnarray}
\lefteqn{A_{2a} \equiv \frac14 ( -i\lambda ) (-\frac{i}{2}
\lambda^2 )  \int \!\! d^Dy \sqrt{-g(y)} \, g^{\alpha\beta}(y) 
i\Delta(x;y) } \nonumber \\
& & \hspace{2.5cm} \times \!\! \int \!\! d^Dz \sqrt{-g(z)} \, g^{\mu\nu}(z)
\, \partial^y _{\alpha} \partial^z _{\mu} i\Delta(y;z) \, \partial^y _{\beta} 
\partial^z _{\nu} i\Delta(y;z) \, i\Delta(z;z) \; . \qquad \label{2adef}
\end{eqnarray}
To conserve space in subsequent expressions we will use the Latin letters as superscripts or subscripts to indicate 
the spacetime arguments of the metric, for example, $\sqrt{g_y} \equiv 
\sqrt{-g(y)}$ and ${g}_z ^{\mu\nu} \equiv g^{\mu\nu}(z)$. Also, we use $\partial^y _\mu \equiv \frac{\partial}{\partial y^{\mu}}$ for partial derivatives. We reduce $A_{2a}$, 
and all the other contributions, by partially integrating derivatives and then 
using either the propagator equation (\ref{propeqn}) or else one of the 
coincidence limits (\ref{propID1}-\ref{propID2}). The first step is to partially 
integrate the factor of $\partial^{y}_{\alpha}$ on the last line of (\ref{2adef}),
\begin{eqnarray}
\lefteqn{A_{2a} = \frac{\lambda^3}{8} \! \int \!\! d^Dy i\Delta(x;y) \!\! 
\int \!\! d^Dz \! \sqrt{-g_z} {g}_z ^{\mu\nu} \partial^z _{\mu} i\Delta(y;z)
\partial^z _{\nu} \mathcal{D}_y i\Delta(y;z) i\Delta(z;z) } \nonumber \\
& & \hspace{-0.5cm} + \frac{\lambda^3}{8} \! \int \!\! d^Dy \! \sqrt{-g_y}   
{g}_y ^{\alpha\beta} \partial^y _{\alpha} i\Delta(x;y) \!\! \int \!\! d^Dz 
\sqrt{-g_z}  {g}_z ^{\mu\nu} \partial^z_{\mu} i\Delta(y;z) \partial^y _{\beta} 
\partial^z_{\nu} i\Delta(y;z) i\Delta(z;z) \; . \qquad \label{2astep1}
\end{eqnarray}
Recall that $\mathcal{D}_y \equiv \partial^y _{\alpha} [\sqrt{-g_y} \, 
{g}_y ^{\alpha\beta} \partial^y _{\beta} ]$. 

We can use the propagator equation (\ref{propeqn}) to replace $\mathcal{D}_y 
i\Delta(y;z)$ on the first line of (\ref{2astep1}) with $i \delta^D(y - z)$.
This permits the integration over $y$ to be performed,
\begin{eqnarray}
\lefteqn{A_{2a1} \equiv \frac{\lambda^3}{8} \! \int \! d^D \! y i\Delta(x;y) \!\! 
\int \!\! d^D z  \sqrt{-g_z} {g_z}^{\mu\nu} \partial^z _{\mu} i\Delta(y;z)
\partial^z_{\nu} i \delta^D(y \!-\! z) i\Delta(z;z) } \nonumber \\
& & \hspace{3.4cm} = \frac{i \lambda^3}{8} \! \int \!\! d^Dz \! \sqrt{-g_z} 
{g_z}^{\mu\nu} \partial^y _{\nu} \Bigl[ i \Delta(x;y) \partial^z_{\mu} 
i\Delta(y;z) \Bigr]_{y=z} \! i\Delta(z;z) \; . \qquad 
\label{2a1step2}
\end{eqnarray}
The differentiated square bracket can be evaluated using 
(\ref{propID1}-\ref{propID2}),
\begin{eqnarray}
\lefteqn{ \partial^y _{\nu} \Bigl[ i \Delta(x;y) \partial^z_{\mu} i\Delta(y;z) 
\Bigr]_{y=z} } \nonumber \\
& & \hspace{3.5cm} = \partial^z_{\nu} i\Delta(x;z) \partial^z _{\mu} 
i\Delta(y;z) \Bigl\vert_{y = z} + i \Delta(x;z) \partial^y _{\nu}
\partial^z _{\mu} i \Delta(y;z) \Bigl\vert_{y = z} \; , \qquad 
\label{2a1step3} \\
& & \hspace{3.5cm} = \partial^z _{\nu} i \Delta(x;z) \!\times\! k H a_z 
\delta^0_{~\mu} + i \Delta(x;z) \!\times\! -\Bigl( \frac{D\!-\!1}{D} \Bigr)
k H^2 g^z _{\mu\nu} \; . \qquad \label{2a1step4}
\end{eqnarray}
Substituting (\ref{2a1step4}) in (\ref{2a1step2}) with $\sqrt{-g_z} \, 
{g}_z ^{\mu\nu} = {a}_z ^{D-2} \eta^{\mu\nu}$, partially integrating, and then
making use of (\ref{propID2}) completes the reduction of $A_{2a1}$,
\begin{eqnarray}
\lefteqn{ A_{2a1} = -\frac{i \lambda^3}{8} k H \! \int \!\! d^Dz {a}_z ^{D-1} 
\partial^z _0 i\Delta(x;z) i\Delta(z;z) } \nonumber \\
& & \hspace{5.5cm} - \frac{i\lambda^3}{8} (D\!-\! 1) k H^2 \! \int \!\! d^Dz 
\sqrt{-g_z} \, i\Delta(x;z) i \Delta(z;z) \; , \qquad \label{2a1step5} \\
& & \hspace{0.2cm} = \frac{i \lambda^3}{8} k H \! \int \!\! d^Dz {a}_z ^{D-1}
i\Delta(x;z) \partial^z _0 i\Delta(z;z) \; , \qquad \label{2a1step6} \\
& & \hspace{0.2cm} = \frac{i \lambda^3}{4} k^2 H^2 \! \int \!\! d^Dz 
\sqrt{-g_z} \, i\Delta(x;z) \longrightarrow \frac{i \lambda^3}{4} k^2 H^2
\!\times\! I_3 \; . \qquad \label{2a1step7}
\end{eqnarray}

The reduction of the term on the last line of (\ref{2astep1}) begins by
exploiting symmetry to write,
\begin{equation}
\sqrt{-g_z} \, {g}_z ^{\mu\nu} \partial^z _{\mu} i\Delta(y;z) \partial^y _{\beta}
\partial^z _{\nu} i\Delta(y;z) = \frac12 \partial^y_{\beta} \Bigl[ \sqrt{-g_z} 
\, {g}_z ^{\mu\nu} \partial^z _{\mu} i\Delta(y;z) \partial^z_{\nu} 
i\Delta(y;z) \Bigr] . \label{2a2step2}
\end{equation}
We then partially integrate the $\partial^y _{\beta}$ and use the propagator
equation (\ref{propeqn}),
\begin{eqnarray}
\lefteqn{ A_{2a2} \equiv -\frac{\lambda^3}{16} \! \int \!\! d^Dy \sqrt{-g_y} \,
\mathcal{D}_y i\Delta(x;y) \! \int \!\! d^Dz \sqrt{-g_z} \, {g_z}^{\mu\nu}
\partial^z_{\mu} i\Delta(y;z) \partial^z_{\nu} i\Delta(y;z) \,
i \Delta(z;z) } \nonumber \\
& & \hspace{4cm} = -\frac{i \lambda^3}{16} \! \int \!\! d^Dz \sqrt{-g_z} \,
\, {g}_z ^{\mu\nu} \partial^z _{\mu} i\Delta(x;z) \partial^z _{\nu} 
i\Delta(x;z) \, i \Delta(z;z) \; . \qquad \label{2a2step3}
\end{eqnarray}
Partially integrating the $\partial^z _{\mu}$ and exploiting relations 
(\ref{propeqn}) and (\ref{propID2}) gives,
\begin{eqnarray}
\lefteqn{A_{2a2} = \frac{i \lambda^3}{16} \! \int \!\! d^Dz \, i \Delta(x;z)
\mathcal{D}_z i \Delta(x;z) \, i\Delta(z;z) } \nonumber \\
& & \hspace{4.5cm} + \frac{i \lambda^3}{16} \! \int \!\! d^Dz \sqrt{-g_z} \,
{g}_z ^{\mu\nu} i \Delta(x;z) \partial^z _{\nu} i \Delta(x;z) \partial^z _{\mu}
i \Delta(z;z) \; , \qquad \label{2a2step4} \\
& & \hspace{0cm} = -\frac{\lambda^3}{16} \Bigl[ i\Delta(x;x)\Bigr]^2 -
\frac{i \lambda^3}{8} k H \! \int \!\! d^Dz {a}_z ^{D-1} i \Delta(x;z)
\partial^z _0 i\Delta(x;z) \; . \qquad \label{2a2step5}
\end{eqnarray}
Now note that,
\begin{equation}
i \Delta(x;z) \partial^z _0 i\Delta(x;z) = \frac12 \partial^z _0 \Bigl[
i\Delta(x;z)\Bigr]^2 \; , \label{2a2step6}
\end{equation}
and partially integrate to reach the form,
\begin{equation}
A_{2a2} = -\frac{\lambda^3}{16} \Bigl[ i\Delta(x;x)\Bigr]^2 + 
\frac{i \lambda^3}{16} (D\!-\! 1) k H^2 \!\times\! I_1 \; . \label{2a2step7}
\end{equation}
Combining relations (\ref{2a1step7}) and (\ref{2a2step7}) gives the total
for the first diagram,
\begin{equation}
A_{2a} = -\frac{\lambda^3}{16} \Bigl[ i\Delta(x;x)\Bigr]^2 + 
\frac{i \lambda^3}{16} (D\!-\! 1) k H^2 \!\times\! I_1 + \frac{i \lambda^3}{4} 
k^2 H^2 \!\times\! I_3 \; . \label{2afinal}
\end{equation}

For the 2nd diagram of Figure~\ref{alldiag} from the left, it has a symmetry 
factor of $\frac12$, and three 3-point vertices,
\begin{eqnarray}
\lefteqn{ A_{2b} \equiv \frac12  (-i \lambda)^3 \! 
\int \!\! d^Dy \sqrt{-g_y} \, {g}_y^{\alpha\beta} i\Delta(x;y) \!\! \int \!\!
d^Dz \sqrt{-g_z} \, {g}_z ^{\mu\nu} \partial^y _{\alpha} \partial^z _{\mu}
i \Delta(y;z) } \nonumber \\
& & \hspace{4cm} \times \!\! \int \!\! d^Dw \sqrt{-g_w} \, g_w^{\rho\sigma}
\partial^y _{\beta} \partial^w _{\rho} i\Delta(y;w) \!\times\! \partial^z _{\nu} 
\partial^w _{\sigma} i\Delta(z;w) \, i \Delta(z;w) \; . \qquad \label{2bdef}
\end{eqnarray}
Applying the same sorts of reductions as for $A_{2a}$ we at length reach 
the form,
\begin{equation}
A_{2b} = \frac{3 \lambda^3}{16} \Bigl[ i\Delta(x;x)\Bigr]^2 - 
\frac{i \lambda^3}{8} (D \!-\! 1) k H^2 \!\times\! I_1 - \frac{3 i \lambda^3}{4}
k^2 H^2 \!\times\! I_3 \; . \label{2bfinal}
\end{equation}
Now, we have the 3rd diagram of Figure~\ref{alldiag} next. It has a symmetry 
factor of $\frac12$ and combines a 3-point vertex with an insertion of the 
2-point counterterm (\ref{B2cterms}),
\begin{eqnarray}
\lefteqn{A_{2c} \equiv \frac12 ( -i\lambda )( -i C_{2B^2} ) \! \int \!\! d^Dy \sqrt{-g_y} \, {g}_y ^{\alpha\beta} i \Delta(x;y) }
\nonumber \\
& & \hspace{5.5cm} \times \!\! \int \!\! d^Dz \sqrt{-g_z} \, {g}_z^{\mu\nu} 
\partial^y _{\alpha} \partial^z _{\mu} i\Delta(y;z) \partial^y _{\beta} 
\partial^z _{\nu} i\Delta(y;z) \, R \; . \qquad \label{2cdef} 
\end{eqnarray}
Note that the counterterm proportional to $C_{1B^2}$ vanishes in dimensional
regularization. Note also that the Ricci scalar $R = D (D-1) H^2$ is constant
on de Sitter background. After some straightforward reductions we reach the
form,
\begin{equation}
A_{2c} = -\frac{\lambda C_{2B^2}}{4} \, R \, i\Delta(x;x) \; . \label{2cfinal}
\end{equation}

The 4th diagram of 
Figure~\ref{alldiag} has a symmetry factor of $\frac12$ and combines 
4-point and 3-point vertices,
\begin{eqnarray}
\lefteqn{ A_{2d} \equiv \frac12 ( -\frac{i \lambda^2}{2}) ( 
-i \lambda) \int \!\! d^Dy \sqrt{-g_y} \, {g}_y^{\alpha\beta} 
i \Delta(x;y) } \nonumber \\
& & \hspace{4cm} \times\! \int \!\! d^Dz \sqrt{-g_z} \, {g}_z ^{\mu\nu}
\partial^y _{\alpha} \partial^z _{\mu} i \Delta(y;z) \partial^y _{\beta} 
\partial^z_{\nu} i\Delta(y;z) \, i \Delta(y;z) \; . \qquad \label{2ddef}
\end{eqnarray}
After many reductions we find,
\begin{equation}
A_{2d} = -\frac{i \lambda^3}{4} (D\!-\!1) k H^2 \!\times\! I_2 - 
\frac{i \lambda^3}{8} k^2 H^2 \!\times\! I_3 \; . \label{2dfinal} 
\end{equation}
The 5th of the diagrams in Figure~\ref{alldiag} has a symmetry factor of $\frac14$ and also involves
4-point and 3-point vertices,
\begin{eqnarray}
\lefteqn{ A_{2e} \equiv \frac14  (-\frac{i \lambda^2}{2}) 
(-i \lambda ) \int \!\! d^Dy \sqrt{-g_y} \, {g}_y^{\alpha\beta} 
i \Delta(x;y) \!\times\! \partial^y _{\alpha} \partial^z _{\beta} i\Delta(y;z) 
\Bigr\vert_{z=y} } \nonumber \\
& & \hspace{4cm} \times\! \int \!\! d^D w \sqrt{-g_w} \, {g}_w ^{\mu\nu} 
i \Delta(y;z) \!\times\! \partial^w _{\mu} \partial^v _{\nu} i \Delta(w;v) 
\Bigl\vert_{v = w} \; . \qquad \label{2edef}
\end{eqnarray}
Two applications of (\ref{propID2}) reduce this diagram to,
\begin{equation}
A_{2e} = -\frac{\lambda^3}{8} (D\!-\!1)^2 k^2 H^4 \!\times\! I_4 \; . \label{2efinal}
\end{equation}

The penultimate diagram from Figure~\ref{alldiag} has a symmetry factor of $\frac14$ and involves three
3-point vertices,
\begin{eqnarray}
\lefteqn{ A_{2f} \equiv \frac14  (-i\lambda)^3 \! \int \!\!
d^Dy \sqrt{-g_y} \, {g}_y ^{\alpha\beta} i\Delta(x;y) \! \int \!\! d^Dz \sqrt{-g_z}
\, {g}_z ^{\mu\nu} \partial^y _{\alpha} \partial^z _{\mu} i\Delta(y;z) \partial^y _{\beta}
\partial^z _{\nu} i\Delta(y;z) } \nonumber \\
& & \hspace{4cm} \times\! \int \!\! d^D w \sqrt{-g_w} \, {g}_w ^{\rho\sigma}
i\Delta(z;w) \!\times\! \partial^w _{\rho} \partial^v _{\sigma} 
i\Delta(w;v) \Bigl\vert_{v = w} \; . \qquad \label{2fdef}
\end{eqnarray}
A long series of reductions yields the form,
\begin{equation}
A_{2f} = \frac{i \lambda^3}{16} (D\!-\!1) k H^2 \!\times\! I_1 -\frac{i \lambda^3}{8}
(D\!-\!1) k H^2 \!\times\! i\Delta(x;x) \!\times\! I_3 - \frac{\lambda^3}{4} (D\!-\!1)
k^2 H^3 \!\times\! I_5 \; . \label{2ffinal}
\end{equation}
Lastly, we have the final diagram of Figure~\ref{alldiag}.
It has a symmetry factor of $\frac12$ and involves the 3-point counterterm 
(\ref{AB2cterms}),
\begin{eqnarray}
\lefteqn{A_{2g} = \frac12 ( -i C_{1 AB^2}) \! \int \!\! d^D y
\, \mathcal{D}_y i\Delta(x;y) \!\times\! {g}_y ^{\alpha\beta} \partial^y _{\alpha}
\partial^z _{\beta} i\Delta(y;z) \Bigl\vert_{z=y} + 0 + 0 } \nonumber \\
& & \hspace{3cm} + \frac12 ( -i C_{4 AB^2}) \! \int \!\! d^Dy 
\sqrt{-g_y} {g}_y ^{\alpha\beta} i\Delta(x;y) \!\times\! R \partial^y _{\alpha} 
\partial^z _{\beta} i\Delta(y;z) \Bigl\vert_{z=y} \; . \qquad \label{2gdef}
\end{eqnarray}
Application of (\ref{propeqn}) and (\ref{propID2}) gives,
\begin{equation}
A_{2g} = -\frac{C_{1 AB^2}}{2} (D\!-\!1) k H^2 + 
\frac{i C_{4 AB^2}}{2} (D\!-\!1) k H^2 \!\times\! I_3 \; . \label{2gfinal}
\end{equation}

We now extract the leading logarithm contributions from each diagram (\ref{2afinal}),
(\ref{2bfinal}), (\ref{2cfinal}), (\ref{2dfinal}), (\ref{2efinal}), (\ref{2ffinal})
and (\ref{2gfinal}). Of course the coincidence limit of the propagator and its 
square are,
\begin{equation}
i\Delta(x;x) \longrightarrow \frac{H^2}{4 \pi^2} \!\times\! \ln(a) \qquad , \qquad
\Bigl[ i\Delta(x;x)\Bigr]^2 \longrightarrow \frac{H^4}{16 \pi^4} \!\times\! \ln^2(a)
\; . \label{coincprops}
\end{equation}
In the Appendix we show that the leading logarithm contributions from the five 
integrals (\ref{I1def}-\ref{I45def}) are,
\begin{eqnarray}
I_1 \longrightarrow -\frac{i}{12 \pi^2} \!\times\! \ln^2(a) & , & \label{I1final} \\
I_2 \longrightarrow -\frac{i}{24 \pi^2} \!\times\! \ln^2(a) & \qquad , \qquad &
I_3 \longrightarrow -\frac{i}{3 H^2} \!\times\! \ln(a) \; , \qquad 
\label{I23final} \\
I_4 \longrightarrow -\frac1{18 H^4} \!\times\! \ln^2(a) & \qquad , \qquad &
I_5 \longrightarrow -\frac1{9 H^3} \!\times\! \ln(a) \; . \qquad \label{I45final}
\end{eqnarray}
Using relations (\ref{coincprops}) and (\ref{I1final}-\ref{I45final}) we can
express the leading logarithm contribution from each of the seven diagrams in
Figure~\ref{alldiag} as a number times the stochastic prediction of $S = \lambda^3
H^4 \ln^2(a)/2^{10} \pi^4$,
\begin{eqnarray}
A_{2a} \longrightarrow -2 \!\times\! S & \qquad , \qquad & A_{2b} \longrightarrow
+8 \!\times\! S \; , \qquad \label{2abS} \\
A_{2c} \longrightarrow +0 \!\times\! S & \qquad , \qquad & A_{2d} \longrightarrow
-4 \!\times\! S \; , \qquad \label{2cdS} \\
A_{2e} \longrightarrow +1 \!\times\! S & \qquad , \qquad & A_{2f} \longrightarrow
-2 \!\times\! S \; , \qquad \label{2efS} \\
A_{2g} \longrightarrow +0 \!\times\! S & \qquad . \qquad & \label{2gS}
\end{eqnarray}
Even if we discount the counterterm diagrams $A_{2c}$ and $A_{2g}$, which cannot
contribute at leading logarithm order, the fact that the remaining five diagrams 
contrive to add up to $+1 \times S$ represents an impressive confirmation of the 
stochastic prediction (\ref{stochastic}).

\subsection{Implications}

The obvious implication is that the stochastic prediction of section 2.3 is correct
to all orders. Because this prediction consists of a ``classical'' evolution 
(\ref{Aclass}), which is accelerated by stochastic jitter, we know that the background
rolls down its effective potential (\ref{Veff}) for all time, reaching arbitrarily
large values. Note that this model provides an explicit contradiction to the view
which is sometimes expressed that large inflationary logarithms must always sum up to 
produce a static, de Sitter invariant result at late times. 

\section{The 1-Loop Beta Function}

\begin{figure}[H]
\centering
\includegraphics[width=16cm]{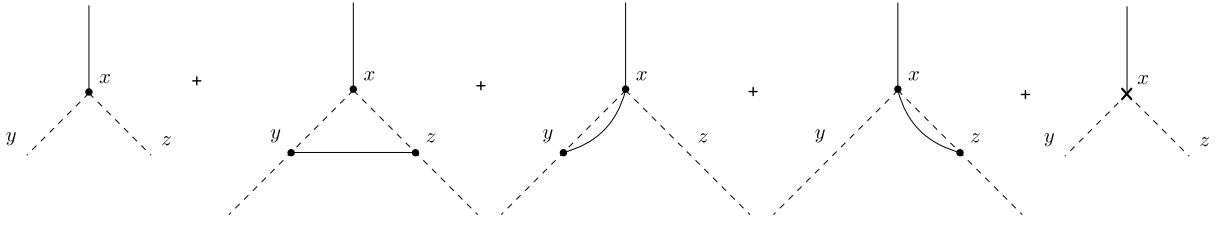}
\caption{\footnotesize Diagrams contributing to the $AB^2$ vertex $-i V(x;y;z)$
at tree and 1-loop orders. Recall that $A$ lines are solid whereas $B$ lines are 
dashed.}
\label{allbetadiag}
\end{figure}

The purpose of this section is to compute the 1-loop beta function (\ref{beta}).
This requires renormalizing the $AB^2$ vertex $-i V(x;y;z)$, the diagrams for 
which are shown in Figure~\ref{allbetadiag}. 
\noindent We first use partial integration to reduce each diagram as much as 
possible. Then derivatives are extracted from products of propagators so as to 
make the result integrable in $D=4$ dimensions,
\begin{equation}
\int \!\! d^Dy \, F(y) \int \!\! d^Dz \, G(z) \!\times\! -i V(x;y;z) \; ,
\label{Vint}
\end{equation}
where $F(y)$ and $G(z)$ are smooth functions. Actually obtaining a finite result
requires adding zero in the form,
\begin{equation}
\partial^2 \Bigl[ \frac1{\Delta x^{D-2}} \Bigr] - 
\frac{4 \pi^{\frac{D}2} i \delta^D(\Delta x)}{\Gamma(\frac{D}2 \!-\! 1)} = 0 
\; , \label{zero}
\end{equation}
where $\Delta x^{\mu}$ is either $(x - y)^{\mu}$ or $(x - z)^{\mu}$. This 
localizes the divergences so that they can be removed by the counterterm 
(\ref{AB2cterms}). We then compute the 1-loop beta function, and employ the
renormalization group to solve the Callan-Symanzik equation to all orders.

\subsection{Partial Integration Reductions}

We start with the first diagrams of Figure~\ref{allbetadiag}.
It is the tree order vertex,
\begin{equation}
-i V_{0}(x;y;z) = -i \lambda \sqrt{-g_x} \, g_x^{\mu\nu} \partial_{\mu} 
\delta^D(x \!-\! y) \partial_{\nu} \delta^D(x \!-\! z) \; . \label{V0}
\end{equation}
The 2nd diagram of Figure~\ref{allbetadiag} gives the most complicated of the 1-loop contributions,
\begin{eqnarray}
\lefteqn{-i V_{1a}(x;y;z) = (-i\lambda)^3  
\Bigl(-\frac{\partial}{\partial y^{\sigma}} \Bigr) \Bigl( -\frac{\partial}{\partial z^{\beta}} \Bigr)
 \Biggl[ \sqrt{-g_x} \, g_x^{\mu\nu} \sqrt{-g_y} \, g_y^{\rho\sigma} 
\partial^x_{\mu} \partial^y_{\rho} i\Delta(x;y) } \nonumber \\
& & \hspace{8.8cm} \times\! \sqrt{-g_z} \, g_z^{\alpha\beta} \partial^x_{\nu}
\partial^z_{\alpha} i\Delta(x;z) \, i\Delta(y;z) \Biggr] . \qquad \label{V1a}
\end{eqnarray}
Note that the derivatives on the external $y^{\sigma}$ and $z^{\beta}$ legs
are partially integrated to act back on the entire diagram. By applying the
same partial integration techniques we used in the previous section, and recalling
that $\mathcal{D} \equiv \partial_{\mu} [\sqrt{-g} \, g^{\mu\nu} \partial_{\nu}]$,
one can
reduce $-i V_{1a}$ to the form,
\begin{eqnarray}
\lefteqn{ -i V_{1a}(x;y;z) = \frac{i \lambda^3}{2} \mathcal{D}_x \partial^y_{\sigma}
\partial^z_{\beta} \Biggl\{ \sqrt{-g_y} \, g_y^{\rho\sigma} \partial^y_{\rho}
i\Delta(x;y) \sqrt{-g_z} \, g_z^{\alpha\beta} \partial^z_{\alpha} i\Delta(x;z) \,
i\Delta(y;z) \Biggr\} } \nonumber \\
& & \hspace{6cm} + \frac{\lambda^3}{4} \mathcal{D}_y \mathcal{D}_z \Biggl\{ 
\Bigl[ i\Delta(y;z)\Bigr]^2 \Bigl[ \delta^D(x \!-\! y) + \delta^D(x \!-\! z)\Bigr]
\Biggr\} \nonumber \\
& & \hspace{1cm} - \frac{\lambda^3}{2} \partial^y_{\sigma} \partial^z_{\beta}
\Biggl\{ \sqrt{-g_y} \, g_y^{\rho\sigma} \partial^y_{\rho} i\Delta(y;z)
\sqrt{-g_z} \, g_z^{\alpha\beta} \partial^z_{\alpha} i\Delta(y;z) \Bigl[
\delta^D(x \!-\! y) + \delta^D(x \!-\! z)\Bigr] \Biggr\} . \qquad \label{V1asemi}
\end{eqnarray}

The 3rd and 4th diagrams of Figure~\ref{allbetadiag} are the next ones that we reduce. The 3rd diagram has a 4-point vertex at $x$ and a 
3-point vertex at $y$, with the external derivative with respect to $y^{\sigma}$ 
partially integrated back on the whole diagram,
\begin{eqnarray}
\lefteqn{ -i V_{1b}(x;y;z) = -i\lambda \Bigl( -\frac{i \lambda^2}{2} \Bigr) \Bigl(
-\frac{\partial}{\partial y^{\sigma}} \Bigr) \Biggl[ \sqrt{-g_x} \, g_x^{\mu\nu} 
\sqrt{-g_y} \, g_y^{\rho\sigma} \partial^x_{\mu} \partial^y_{\rho} i\Delta(x;y) }
\nonumber \\
& & \hspace{10.8cm} \times i\Delta(x;y) \partial_{\nu} \delta^D(x \!-\! z) 
\Biggr] . \qquad \label{V1b}
\end{eqnarray}
Partial integration reduces this to,
\begin{eqnarray}
\lefteqn{-i V_{1b}(x;y;z) = \frac{\lambda^3}{4} \mathcal{D}_y \Biggl\{ \sqrt{-g_x} \,
g_x^{\mu\nu} \partial^x_{\mu} \Bigl[ i\Delta(x;y)\Bigr]^2 \partial_{\nu}
\delta^D(x \!-\!z) \Biggr\} } \nonumber \\
& & \hspace{3.5cm} - \frac{\lambda^3}{2} \partial^y_{\sigma} \Biggl\{ \sqrt{-g_x} \,
g_x^{\mu\nu} \partial^x_{\mu} i\Delta(x;y) \sqrt{-g_y} \, g_y^{\rho\sigma}
\partial^y_{\rho} i\Delta(x;y) \partial_{\nu} \delta^D(x \!-\! z) \Biggr\} . \qquad
\label{V1bsemi}
\end{eqnarray}
The 4th diagram has a 4-point vertex at $x$ and a 3-point vertex at $z$, 
with the external derivative with respect to $z^{\beta}$ acted back on everything,
\begin{eqnarray}
\lefteqn{ -i V_{1c}(x;y;z) = -i\lambda \Bigl( -\frac{i \lambda^2}{2} \Bigr) \Bigl(
-\frac{\partial}{\partial z^{\beta}} \Bigr) \Biggl[ \sqrt{-g_x} \, g_x^{\mu\nu} 
\sqrt{-g_z} \, g_z^{\alpha\beta} \partial^x_{\mu} \partial^z_{\alpha} i\Delta(x;z) }
\nonumber \\
& & \hspace{10.8cm} \times i\Delta(x;z) \partial_{\nu} \delta^D(x \!-\! y) 
\Biggr] . \qquad \label{V1c}
\end{eqnarray}
The same partial integrations give,
\begin{eqnarray}
\lefteqn{-i V_{1c}(x;y;z) = \frac{\lambda^3}{4} \mathcal{D}_z \Biggl\{ \sqrt{-g_x} \,
g_x^{\mu\nu} \partial^x_{\mu} \Bigl[ i\Delta(x;z)\Bigr]^2 \partial_{\nu}
\delta^D(x \!-\!y) \Biggr\} } \nonumber \\
& & \hspace{3.5cm} - \frac{\lambda^3}{2} \partial^z_{\beta} \Biggl\{ \sqrt{-g_x} \,
g_x^{\mu\nu} \partial^x_{\mu} i\Delta(x;z) \sqrt{-g_z} \, g_z^{\alpha\beta}
\partial^z_{\alpha} i\Delta(x;z) \partial_{\nu} \delta^D(x \!-\! y) \Biggr\} . \qquad
\label{V1csemi}
\end{eqnarray}

It turns out that the last lines of expressions (\ref{V1asemi}), (\ref{V1bsemi}) 
and (\ref{V1csemi}) cancel. To see this, use the delta function on the last line of 
(\ref{V1csemi}) to flip the $x$ derivative to a $y$ derivative, then extract the
derivative and use the delta function to convert $x$ to $y$,
\begin{eqnarray}
\lefteqn{ - \frac{\lambda^3}{2} \partial^z_{\beta} \Biggl\{ \sqrt{-g_x} \,
g_x^{\mu\nu} \partial^x_{\mu} i\Delta(x;z) \sqrt{-g_z} \, g_z^{\alpha\beta}
\partial^z_{\alpha} i\Delta(x;z) \partial_{\nu} \delta^D(x \!-\! y) \Biggr\} }
\nonumber \\
& & \hspace{3cm} = \frac{\lambda^3}{2} \partial^y_{\nu} \partial^z_{\beta}
\Biggl\{ \sqrt{-g_x} \, g_x^{\mu\nu} \partial^x_{\mu} i\Delta(x;z) \sqrt{-g_z} 
\, g_z^{\alpha\beta} \partial^z_{\alpha} i\Delta(x;z) \delta^D(x \!-\! y) 
\Biggr\} \; , \qquad \\
& & \hspace{3cm} = \frac{\lambda^3}{2} \partial^y_{\sigma} \partial^z_{\beta}
\Biggl\{ \sqrt{-g_y} \, g_y^{\rho\sigma} \partial^y_{\rho} i\Delta(y;z) 
\sqrt{-g_z} \, g_z^{\alpha\beta} \partial^z_{\alpha} i\Delta(y;z) 
\delta^D(x \!-\! y) \Biggr\} \; . \qquad
\label{cancellation}
\end{eqnarray}
Similar manipulations result in partial cancellations between the first lines 
of (\ref{V1bsemi}) and (\ref{V1csemi}) and the second line of (\ref{V1asemi}).
The resulting sum of all 1-loop primitive contributions to the vertex is,
\begin{eqnarray}
\lefteqn{ -i V_{1 \rm prim} = \frac{i\lambda^3}{2} \mathcal{D}_x \partial_y^{\rho}
\partial_z^{\alpha} \Bigl\{ (a_y a_z)^{D-2} i\Delta(y;z) \partial^y_{\rho}
i\Delta(x;y) \partial^z_{\alpha} i\Delta(x;z) \Bigr\} } \nonumber \\
& & \hspace{0cm} + \frac{\lambda^3}{4} \mathcal{D}_y \partial_z^{\alpha} 
\Bigl\{ a_z^{D-2} \Bigl[ i\Delta(y;z)\Bigr]^2 \partial^z_{\alpha} 
\delta^D(x \!-\! z) \Bigr\} + \frac{\lambda^3}{4} \mathcal{D}_z \partial_y^{\rho} 
\Bigl\{ a_y^{D-2} \Bigl[ i\Delta(y;z)\Bigr]^2 \partial^y_{\sigma} 
\delta^D(x \!-\! y) \Bigr\} \; . \qquad \label{V1primitive}
\end{eqnarray}

\subsection{Extracting Divergences}

The three curly bracketed expressions of the primitive contribution 
(\ref{V1primitive}) are each logarithmically divergent. This means that
divergences derive entirely from the first term in the expansion 
(\ref{propagator}-\ref{Fprime}) of the propagator,
\begin{equation}
i\Delta(y;z) = \frac{\Gamma(\frac{D}2 \!-\! 1)}{4 \pi^{\frac{D}{2}}}
\frac1{[a_y a_z (y \!-\! z)^2]^{\frac{D}2 - 1}} + \dots \label{propexp}
\end{equation}
For example, the second and third terms of (\ref{V1primitive}) involve 
the propagator squared,
\begin{eqnarray}
\lefteqn{\Bigl[ i\Delta(y;z)\Bigr]^2 = \frac{\Gamma^2(\frac{D}2 \!-\!1)}{
16 \pi^D} \frac1{[ a_y a_z (y \!-\! z)^2]^{D-2}} + {\rm UV\ Finite} \; ,
} \\
& & \hspace{-0.2cm} \longrightarrow \frac{\Gamma(\frac{D}2 \!-\!1)}{4 \pi^{\frac{D}2}}
\frac{\mu^{D-4}}{2 (D\!-\!3) (D\!-\!4)} \frac{i \delta^D(y \!-\! z)}{
(a_y a_z)^{D-2}} - \frac{\partial^2}{64 \pi^4 (a_y a_z)^2} \Biggl[
\frac{\ln[\mu^2 (y \!-\! z)^2]}{(y \!-\! z)^2} \Biggr] + {\rm UV\ Finite} \; .
\qquad \label{propsq}
\end{eqnarray}

The key to extracting the divergence of the first term in (\ref{V1primitive})
is noting that it must be proportional to the Minkowski metric,
\begin{eqnarray}
\lefteqn{ (a_y a_z)^{D-2} i\Delta(y;z) \partial^y_{\rho} i\Delta(x;y) 
\partial^z_{\alpha} i\Delta(x;z) } \nonumber \\
& & \hspace{1.4cm} = \frac{\Gamma^3(\frac{D}2 \!-\! 1)}{64 \pi^{\frac{3D}{2}}}
\frac1{a_x^{D-2} (y \!-\! z)^{D-2}} \frac{\partial}{\partial y^{\rho}}
\Bigl[ \frac1{(x \!-\! y)^{D-2}}\Bigr] \frac{\partial}{\partial z^{\alpha}}
\Bigl[ \frac1{(x \!-\! z)^{D-2}}\Bigr] + {\rm UV\ Finite} \; , \qquad \\
& & \hspace{1.4cm} = \frac{\Gamma^3(\frac{D}2 \!-\! 1)}{64 \pi^{\frac{3D}{2}}}
\frac{\frac1{D}\eta_{\rho\alpha} \eta^{\mu\nu}}{a_x^{D-2} (y \!-\! z)^{D-2}} 
\frac{\partial}{\partial x^{\mu}} \Bigl[ \frac1{(x \!-\! y)^{D-2}}\Bigr] 
\frac{\partial}{\partial x^{\nu}} \Bigl[ \frac1{(x \!-\! z)^{D-2}}\Bigr] + 
{\rm UV\ Finite} \; . \qquad \label{3props}
\end{eqnarray}
It is simple to extract a d'Alembertian from the contracted derivatives,
\begin{eqnarray}
\lefteqn{ \partial^{\mu} \Bigl[ \frac1{(x \!-\! y)^{D-2}} \Bigr] \partial_{\mu}
\Bigl[ \frac1{(x \!-\! z)^{D-2}} \Bigr] = \frac{\partial^2}{2} \Bigl[ 
\frac1{(x \!-\! y)^{D-2}} \frac1{(x \!-\! z)^{D-2}} \Bigr] } \nonumber \\
& & \hspace{4.5cm} - \frac{\partial^2}{2} \Bigl[ \frac1{(x \!-\! y)^{D-2}} \Bigr]
\frac1{(x \!-\! z)^{D-2}} - \frac1{(x \!-\! y)^{D-2}} \frac{\partial^2}{2}
\Bigl[ \frac1{(x \!-\! z)^{D-2}} \Bigr] \; , \qquad \\
& & \hspace{1cm} = \frac{\partial^2}{2} \Bigl[ \frac1{(x \!-\! y)^{D-2}} 
\frac1{(x \!-\! z)^{D-2}} \Bigr] - \frac{2 \pi^{\frac{D}2} i \delta^D(x \!-\!y)}{
\Gamma(\frac{D}2\!-\! 1) (x \!-\! z)^{D-2}} - \frac{2 \pi^{\frac{D}2}
i \delta^D(x \!-\! z)}{\Gamma(\frac{D}2 \!-\! 1)(x \!-\! y)^{D-2}} \; . \qquad
\label{contracted} 
\end{eqnarray}
Substituting (\ref{contracted}) into (\ref{3props}) allows us to isolate and
localize the divergence,
\begin{eqnarray}
\lefteqn{ (a_y a_z)^{D-2} i\Delta(y;z) \partial^y_{\rho} i\Delta(x;y) 
\partial^z_{\alpha} i\Delta(x;z) } \nonumber \\
& & \hspace{3.5cm} = -\frac{\Gamma^2(\frac{D}2 \!-\! 1)}{32 \pi^D} \frac{\frac1{D}
\eta_{\rho\alpha}}{a_x^{D-2}} \Biggl\{ \frac{i \delta^D(x \!-\! y)}{(x \!-\! 
z)^{2D-4}} + \frac{i \delta^D(x \!-\! z)}{(x \!-\! y)^{2D-4}} \Biggr\} +
{\rm UV\ Finite} \; , \qquad \\
& & \hspace{3.5cm} = \frac{\Gamma(\frac{D}2 \!-\! 1)}{8 \pi^{\frac{D}2}} 
\frac{\mu^{D-4} \eta_{\rho\alpha}}{D (D\!-\!3) (D\!-\!4)} \frac{\delta^D(x \!-\!y)
\delta^D(x \!-\! z)}{a_x^{D-2}} + {\rm UV\ Finite} \; . \qquad \label{3propsdiv}
\end{eqnarray}

The total divergent part of the primitive contribution to $-i V_1(x;y;z)$ comes 
from substituting expressions (\ref{propsq}) and (\ref{3propsdiv}) in
(\ref{V1primitive}),
\begin{eqnarray}
\lefteqn{-i V_{1 \rm div} = \frac{i\lambda^3}{2} \mathcal{D}_x \partial_y^{\rho}
\partial_z^{\alpha} \Biggl\{ \frac{\Gamma(\frac{D}2 \!-\! 1)}{8 \pi^{\frac{D}2}}
\frac{\mu^{D-4} \eta_{\rho\alpha}}{D (D \!-\! 3) (D\!-\! 4)} 
\frac{\delta^D(x \!-\!y) \delta^D(x \!-\! z)}{a_x^{D-2}} \Biggr\} } \nonumber \\
& & \hspace{1.5cm} + \frac{\lambda^3}{4} \mathcal{D}_y \partial_z^{\alpha}
\Biggl\{ a_z^{D-2} \frac{\Gamma(\frac{D}2 \!-\! 1)}{4 \pi^{\frac{D}2}}
\frac{\mu^{D-4}}{2 (D\!-\!3) (D\!-\! 4)} \frac{i \delta^D(y \!-\! z)}{(a_y a_z)^{D-2}}
\, \partial^z_{\alpha} \delta^D(x \!-\! z) \Biggr\} \nonumber \\
& & \hspace{3cm} + \frac{\lambda^3}{4} \mathcal{D}_z \partial_y^{\rho}
\Biggl\{ a_y^{D-2} \frac{\Gamma(\frac{D}2 \!-\! 1)}{4 \pi^{\frac{D}2}}
\frac{\mu^{D-4}}{2 (D\!-\!3) (D\!-\! 4)} \frac{i \delta^D(y \!-\! z)}{(a_y a_z)^{D-2}}
\, \partial^y_{\rho} \delta^D(x \!-\! y) \Biggr\} . \qquad 
\end{eqnarray}
This expression can be simplified considerably by using the delta functions to
reflect derivatives and to evaluate the free scale factors at $x$. The final 
result is,
\begin{eqnarray}
\lefteqn{ -i V_{1 \rm div} = \frac{i \lambda^3 \mu^{D-4}}{16 \pi^{\frac{D}2}}
\frac{\Gamma(\frac{D}2 \!-\! 1)}{2 (D\!-\!3) (D\!-\!4)} \Biggl\{ \frac{2}{D}
\mathcal{D}_x \Biggl[ \frac{\partial^{\mu} \delta^D(x \!-\! y) \partial_{\mu}
\delta^D(x \!-\! z)}{a_x^{D-2}} \Biggr] } \nonumber \\
& & \hspace{3.3cm}  + \mathcal{D}_y \partial_x^{\mu} \Biggl[ 
\frac{\delta^D(x \!-\!y) \partial_{\mu} \delta^D(x \!-\! z)}{a_x^{D-2}} \Biggr]
+ \mathcal{D}_z \partial_x^{\mu} \Biggl[ \frac{\partial_{\mu} \delta^D(x \!-\!y) 
\delta^D(x \!-\! z)}{a_x^{D-2}} \Biggr] \Biggr\} . \qquad \label{V1div}
\end{eqnarray}

\subsection{The Beta Function}

\begin{figure}[H]
\centering
\includegraphics[width=3cm]{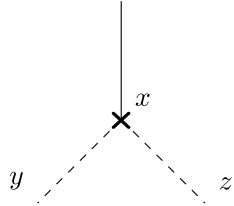}
\caption{\footnotesize The final diagram on Figure~\ref{allbetadiag} which represents
an insertion of the counterterm (\ref{AB2cterms}).}
\label{diagbetacounter}
\end{figure}

The primitive divergence (\ref{V1div}) is canceled using the general 1-loop 
counterterm (\ref{AB2cterms}). Varying the counter-action gives,
\begin{eqnarray}
\lefteqn{\frac{i\delta^3 S_{AB^2}[A,B]}{\delta A(x) \delta B(y) \delta B(z)} = 
-i C_{1 AB^2} \mathcal{D}_x \Biggl[ \frac{\partial^{\mu} \delta^D(x \!-\! y) 
\partial_{\mu} \delta^D(x \!-\! z)}{a_x^2} \Biggr] } \nonumber \\
& & \hspace{0.7cm} 
+ i C_{2 AB^2} \mathcal{D}_y \partial_x^{\mu} \Biggl[
\frac{\delta^D(x \!-\! y) \partial_{\mu} \delta^D(x \!-\! z)}{a_x^2} \Biggr]
+ i C_{2 AB^2} \mathcal{D}_z \partial_x^{\mu} \Biggl[
\frac{\partial_{\mu} \delta^D(x \!-\! y) \delta^D(x \!-\! z)}{a_x^2} \Biggr]
\nonumber \\
& & \hspace{1.4cm} -i C_{3 AB^2} \mathcal{D}_y \mathcal{D}_z \Biggl[
\frac{\delta^D(x \!-\! y) \delta^D(x \!-\! z)}{a_x^D} \Biggr] - i C_{4 AB^2} R 
a_x^{D-2} \partial^{\mu} \delta^D(x \!-\! y) \partial_{\mu} \delta^D(x \!-\! z)
\; . \qquad \label{cterm} 
\end{eqnarray}
Comparison of expressions (\ref{V1div}) and (\ref{cterm}) shows that the
four coefficients are,
\begin{eqnarray}
C_{1 AB^2} & \!\!\! = \!\!\! & \frac{\lambda^3 \mu^{D-4}}{16 \pi^{\frac{D}2}}
\frac{\Gamma(\frac{D}2 \!-\! 1)}{D (D\!-\!3) (D\!-\!4)} \; , \qquad 
\label{C1} \\ 
C_{2 AB^2} & \!\!\! = \!\!\! & -\frac{\lambda^3 \mu^{D-4}}{16 \pi^{\frac{D}2}}
\frac{\Gamma(\frac{D}2 \!-\! 1)}{2 (D\!-\!3) (D\!-\!4)} \; , \qquad 
\label{C2} \\
C_{3 AB^2} & \!\!\! = \!\!\! & 0 \; , \qquad \label{C3} \\
C_{4 AB^2} & \!\!\! = \!\!\! & 0 \; . \qquad \label{C4}
\end{eqnarray}
Hence there is no curvature-dependent coupling constant renormalization at 
1-loop and the beta function is zero at order $\lambda^3$,
\begin{equation}
\delta \lambda = C_{4 AB^2} \!\times\! R + O(\lambda^5) \qquad \Longrightarrow
\qquad \beta = \frac{\partial \delta \lambda}{\partial \ln(\mu)} = O(\lambda^5)
\; . \label{betafinal}
\end{equation}

\subsection{Implications}

Previous work determined the 1-loop $\gamma$ function for $A$ \cite{Miao:2021gic}.
We now have the 1-loop $\beta$ function, so we can use the Callan-Symanzik equation
to study how the $n$-point Green's functions vary with respect to changes in the
renormalization scale $\mu$,
\begin{equation}
\Biggl[ \mu \frac{\partial}{\partial \mu} + \beta(\lambda) 
\frac{\partial}{\partial \lambda} + n \gamma(\lambda) \Biggr]
G_n\Bigl(x_1;\ldots;x_n;\lambda,\mu\Bigr) = 0 \; . \label{CSeqn}
\end{equation}
(The equation for a 1PI $n$-point function is the same, with the $+n \gamma$
term changed to $-n \gamma$.) One solves (\ref{CSeqn}) by the method of 
characteristics. First, find a running coupling constant 
$\overline{\lambda}(\mu)$ such that,
\begin{equation}
\mu \frac{d \overline{\lambda}}{\partial \mu} = -\beta\Bigl( 
\overline{\lambda}(\mu)\Bigr) \;\; , \;\; \overline{\lambda}(\mu_0)
= \lambda \qquad \Longrightarrow \qquad \beta(\lambda) 
\frac{\partial \overline{\lambda}}{\partial \lambda} = 
\beta(\overline{\lambda}) \; . \label{running}
\end{equation}
Then the Green's function at scale $\mu$ can be expressed in terms of its
value at scale $\mu_0$,
\begin{equation}
G_n\Bigl(x_1;\ldots;x_n;\lambda;\mu\Bigr) = G_n\Bigl(x_1;\ldots;x_n;
\overline{\lambda}(\mu);\mu_0\Bigr) \!\times\!
\exp\Bigl[-n \!\! \int_{\mu_0}^{\mu} \!\! \frac{d\mu'}{\mu'} \gamma\Bigl( 
\overline{\lambda}(\mu') \Bigr)\Bigr] \; . \label{RGsol}
\end{equation}
Our result (\ref{betafinal}) for the curvature-dependent beta function
is zero at order $\lambda^3$. If we make the usual assumption that 1-loop
results dominate, this means that the coupling constant fails to run,
that is, $\overline{\lambda}(\mu) = \lambda$. This implies that the
perturbative factors of $\ln(\mu)$ exponentiate to give a power,
\begin{equation}
G_n\Bigl(x_1;\ldots;x_n;\lambda;\mu\Bigr) = G_n\Bigl(x_1;\ldots;
x_n;\lambda;\mu_0\Bigr) \Bigl[\frac{\mu_0}{\mu}\Bigr]^{n \gamma(\lambda)}
\; . \label{RGresum}
\end{equation}
Scaling is more complicated in cosmology because the Hubble parameter gives
an additional dimensionful parameter, even for massless theories. However, 
if it is valid for the exchange potential to replace $\mu$ by either $r$ or
$1/r$, expression (\ref{RGresum}) provides a highly nontrivial resummation.

\section{Conclusions}

We have revisited the 2-field model (\ref{Lagrangian}) which was analyzed in
a recent study of nonlinear sigma models as a paradigm for how to re-sum large
inflationary logarithms from fields with derivative interactions like those of
quantum gravity \cite{Miao:2021gic}. After reviewing the model in section 2,
we derived a 2-loop result for the evolution of the background in section 3,
and computed the 1-loop beta function in section 4. Our purpose was to check
the stochastic prediction (\ref{stochastic}) for the 2-loop background which
was derived in the previous study \cite{Miao:2021gic}, and to extend the 
renormalization group analysis of the exchange potentials to all orders.

The seven diagrams of Figure~\ref{alldiag} contribute to the 2-loop background, 
and can be reduced to exact analytic expressions (\ref{2afinal}), 
(\ref{2bfinal}), (\ref{2cfinal}), (\ref{2dfinal}), (\ref{2efinal}), 
(\ref{2ffinal}) and (\ref{2gfinal}). When the Schwinger-Keldysh formalism is 
used to extract their leading logarithm contributions (\ref{2abS}-\ref{2gS}), 
the total agrees exactly with the stochastic prediction (\ref{stochastic}). 
Even if one discounts the counterterm insertions (\ref{2dfinal}) and 
(\ref{2gfinal}), this still leaves five intricate, 2-loop diagrams which 
conspire to confirm the stochastic prediction. That must be recognized as a
highly nontrivial check of the stochastic formalism.

The implications of our results were discussed at the end of sections 3 and 4.
Briefly, these are that we can now sum the leading logarithms to all orders as
long as the de Sitter background persists. The scalar background evolves to 
arbitrarily large values, and this evolution continues to late times. The 
factors of $\ln(Hr)$ in the exchange potentials sum up to give powers of $Hr$.

Finally, we again note that it would be interesting to generalize this analysis 
from de Sitter to a general cosmological background which has undergone 
primordial inflation. Good analytic approximations for the key correlators 
(\ref{propID1}-\ref{propID2}) of the stochastic analysis have recently been 
derived \cite{Kasdagli:2023nzj}, so it should be possible to work out what
happens on a general background. Because these correlators transmit the high 
scales of primordial inflation to late times \cite{Kasdagli:2023nzj}, it seems 
as if there may be significant late time effects. It should also be possible to 
generalize the renormalization group analysis to a general cosmological geometry 
because the coefficients of counterterms are universal, independent of the 
background \cite{Barvinsky:1985an}.

\vskip 1cm

\centerline{\bf Acknowledgements}

We thank N. C. Tsamis for discussions on this topic. This work was partially 
supported by NSF grant PHY-2207514 and by the Institute for Fundamental 
Theory at the University of Florida.

\section{Appendix: Schwinger-Keldysh Evaluation of (\ref{I1def}-\ref{I45def})}

Section 3 saw the various 2-loop contributions to the in-out matrix element of
$A(x)$ reduced to five integrals (\ref{I1def}-\ref{I45def}). It turns out that
none of these integrals is real, nor do they even converge owing to the vast
volume of the infinite future. Both embarrassments are due to the fact that
inflationary particle production results in differences between in-out matrix
elements and true expectation values. The Schwinger-Keldysh formalism 
\cite{Schwinger:1960qe,Mahanthappa:1962ex,Bakshi:1962dv,Bakshi:1963bn,
Keldysh:1964ud} is a diagrammatic technique for giving the true expectation
value we want. Fortunately, it is simple to convert in-out matrix elements
such as (\ref{I1def}-\ref{I45def}) to the true expectation values.

There are some excellent review articles on the Schwinger-Keldysh formalism 
\cite{Chou:1984es,Jordan:1986ug,Calzetta:1986ey}. We confine ourselves here 
to summarizing the rules and showing how they allow us to convert in-out
matrix elements to expectation values \cite{Ford:2004wc}. The rules are:
\begin{itemize}
\item{The diagram topology is identical to that of in-out matrix elements.}
\item{The endpoint of every line (internal and external) carries a $\pm$ 
polarity, with $+$ standing for a field which evolves forward in time and 
$-$ corresponding to a field which evolves backwards.}
\item{Vertices (including counterterms) are either all $+$ or all $-$. The 
$+$ vertices are the same as those of in-out matrix elements, whereas $-$ 
vertices are complex conjugated.}
\item{Because propagators have two endpoints, each with its own $\pm$
polarity, there are four propagators. They are related to the Feynman
propagator $i\Delta(x;y)$ as follows,
\begin{eqnarray}
i \Delta_{\scriptscriptstyle ++}(x;y) & = & i\Delta(x;y) \qquad , \qquad 
i\Delta_{\scriptscriptstyle --}(x;y) = [i\Delta(x;y)]^* \; , \qquad \\
i \Delta_{\scriptscriptstyle +-}(x;y) & = & \theta(\eta_x \!-\! \eta_y) 
[i \Delta(x;y)]^* + \theta(\eta_y \!-\! \eta_x) i \Delta(x;y) \; , \qquad \\
i \Delta_{\scriptscriptstyle -+}(x;y) & = & \theta(\eta_x \!-\! \eta_y) 
i \Delta(x;y) + \theta(\eta_y \!-\! \eta_x) [i \Delta(x;y)]^* \; , \qquad
\end{eqnarray}
where $\eta_x$ and $\eta_y$ are the conformal times $x^0$ and $y^0$,
respectively.}
\item{If $\mathcal{D}$ is the kinetic operator, the various propagators
obey,
\begin{equation}
\mathcal{D} i \Delta_{\scriptscriptstyle ++}(x;z) = i \delta^D(x \!-\! z) 
\;\; , \;\; \mathcal{D} i \Delta_{\scriptscriptstyle \pm \mp}(x;z) = 0
\;\; , \;\; \mathcal{D} i \Delta_{\scriptscriptstyle --}(x;z) =
-i \delta^D(x \!-\! z) \; . 
\end{equation}} 
\end{itemize}
\nonumber Two important consequences of these rules are (1) that the 
expectation value of a Hermitian operator such as $A(x)$ is real; and (2)
that the only net contribution from interaction vertices comes from the
past light-cone of external points. 

Because the external $A$ line can be viewed as $+$, we see that the 
Schwinger-Keldysh generalizations of the first three integrals 
(\ref{I1def}-\ref{I23def}) are,
\begin{eqnarray}
I_1(t) & \!\!\! \longrightarrow \!\!\! & \int \!\! d^Dz \sqrt{-g(z)} \, 
\Bigl\{ \Bigl[ i\Delta_{\scriptscriptstyle ++}(x;z) \Bigr]^2 - \Bigl[ 
i\Delta_{\scriptscriptstyle +-}(x;z) \Bigr]^2 \Bigr\} \; , \qquad 
\label{SKI1} \\
I_2(t) & \!\!\! \longrightarrow \!\!\! & \int \!\! d^Dz \sqrt{-g(z)} \,
\Bigl\{ i\Delta_{\scriptscriptstyle ++}(x;z) 
i \Delta_{\scriptscriptstyle ++}(z;z) - i\Delta_{\scriptscriptstyle +-}(x;z) 
i\Delta_{\scriptscriptstyle --}(z;z) \Bigr\} \; , \qquad \label{SKI2} \\
I_3(t) & \!\!\! \longrightarrow \!\!\! & \int \!\! d^Dz \sqrt{-g(z)} \, 
\Bigl\{ i\Delta_{\scriptscriptstyle ++}(x;z) - 
i\Delta_{\scriptscriptstyle +-}(x;z)\Bigr\} \; . \qquad \label{SKI3}
\end{eqnarray}
Because $I_3(t)$ is imaginary, the last two integrals become,
\begin{eqnarray}
I_4(t) & \!\!\! \longrightarrow \!\!\! & \int \!\! d^Dz \sqrt{-g(z)} \, 
\Bigl\{ i\Delta_{\scriptscriptstyle ++}(x;z) - 
i\Delta_{\scriptscriptstyle +-}(x;z)\Bigr\} \!\times\! I_3(t_z) \; , 
\qquad \label{SKI4} \\
I_5(t) & \!\!\! \longrightarrow \!\!\! & \int \!\! d^Dz \sqrt{-g(z)} \, 
\Bigl\{ i\Delta_{\scriptscriptstyle ++}(x;z) - 
i\Delta_{\scriptscriptstyle +-}(x;z)\Bigr\} \!\times\! \dot{I}_3(t_z) \; , 
\qquad \label{SKI5}
\end{eqnarray}
The $++$ and $+-$ propagators in these expressions are obtained from
(\ref{propagator}-\ref{Fprime}) by replacing the Minkowski interval
$\Delta x^2$ in the de Sitter length function $\mathcal{Y}(x;z) = a a' 
H^2 \Delta x^2$ with,
\begin{eqnarray} 
\Delta x^2_{\scriptscriptstyle ++} & \!\!\! = \!\!\! & \Bigl\Vert \vec{x}
\!-\! \vec{z} \Bigr\Vert^2 - \Bigl( \vert \eta \!-\! \eta_z \vert \!-\! i \epsilon
\Bigr)^2 \; , \label{Dx++} \\
\Delta x^2_{\scriptscriptstyle +-} & \!\!\! = \!\!\! & \Bigl\Vert \vec{x}
\!-\! \vec{z} \Bigr\Vert^2 - \Bigl( \eta \!-\! \eta_z \!+\! i \epsilon
\Bigr)^2 \; . \label{Dx+-}
\end{eqnarray}
Note that $\Delta x^2_{\scriptscriptstyle +-}$ and $\Delta x^2_{
\scriptscriptstyle ++}$ agree for $\eta_z > \eta_x$, and are complex conjugates 
for $\eta_z < \eta_x$. This guarantees that the integrals over ${z}^{\mu}$
in expressions (\ref{SKI1}-\ref{SKI5}) only contribute when ${z}^{\mu}$
is in the past light-cone of $x^{\mu}$, and that $++$ and $+-$ differences 
are purely imaginary.

The key to evaluating (\ref{SKI1}-\ref{SKI5}) is understanding that $I_1(t)$
is logarithmically divergent, and the divergent part of $I_2$ is a constant
times $I_3$ through the relation,
\begin{eqnarray}
i\Delta_{\scriptscriptstyle ++}(z;z) = -k \pi \cot\Bigl(\frac{D\pi}{2}\Bigr)
+ \frac{H^2}{4\pi^2} \, \ln(a_z) = i\Delta_{\scriptscriptstyle --}(z;z) \; .
\end{eqnarray}
The remaining integrals are finite. This means that we can exploit the $D=4$ 
form of the propagator,
\begin{eqnarray}
i\Delta(x;z) = \frac{\Gamma(\frac{D}2 \!-\! 1)}{4 \pi^{\frac{D}2}}
\frac1{(a a_z \Delta x^2)^{\frac{D}2 - 1}} - \frac{H^2}{8 \pi^2} \ln\Bigl( 
\frac14 H^2 \Delta x^2\Bigr) + O(D\!-\!4) \; . \label{D4prop}
\end{eqnarray}
The square of the leading term has the reduction,
\begin{eqnarray}
\frac1{\Delta x^{2D-4}_{\scriptscriptstyle ++}} & \!\!\! = \!\!\! &
\frac{\mu^{D-4}}{2 (D\!-\!3) (D\!-\!4)} \frac{4 \pi^{\frac{D}2} i
\delta^D(x \!-\! z)}{\Gamma(\frac{D}2 \!-\! 1)} - \frac{\partial_x^2}{4}
\Biggl[ \frac{\ln(\mu^2 \Delta x^2_{\scriptscriptstyle ++})}{\Delta 
x^2_{\scriptscriptstyle ++}} \Biggr] + O(D\!-\!4) \; , \qquad 
\label{Dx++sq} \\
\frac1{\Delta x^{2D-4}_{\scriptscriptstyle +-}} & \!\!\! = \!\!\! &
- \frac{\partial_x^2}{4} \Biggl[ \frac{\ln(\mu^2 \Delta x^2_{
\scriptscriptstyle +-})}{\Delta x^2_{\scriptscriptstyle +-}} \Biggr] 
+ O(D\!-\!4) \; . \qquad \label{Dx+-sq}
\end{eqnarray}
Derivatives can be extracted from the integration to express inverse powers 
of $\Delta x^2_{\scriptscriptstyle +\pm}$ in terms of logarithms,
\begin{eqnarray}
\frac1{\Delta x^2_{\scriptscriptstyle +\pm}} = \frac{\partial_x^2}{4} 
\Bigl[ \ln(\mu^2 \Delta x^2_{\scriptscriptstyle +\pm})\Bigr] \quad , \quad
\frac{\ln(\mu^2 \Delta x^2_{\scriptscriptstyle +\pm})}{\Delta x^2_{
\scriptscriptstyle +\pm}} = \frac{\partial_x^2}{8} \Bigl[ 
\ln^2(\mu^2 \Delta x^2_{\scriptscriptstyle +\pm}) - 2
\ln(\mu^2 \Delta x^2_{\scriptscriptstyle +\pm}) \Bigr] \; . \label{inverse}
\end{eqnarray}
Differences of these logarithms give,
\begin{eqnarray}
\ln(\mu^2 \Delta x^2_{\scriptscriptstyle ++}) - 
\ln(\mu^2 \Delta x^2_{\scriptscriptstyle +-}) & \!\!\! = \!\!\! &
2 \pi i \, \theta(\Delta \eta \!-\! \Delta r) \; , \qquad \label{log1} \\
\ln^2(\mu^2 \Delta x^2_{\scriptscriptstyle ++}) - 
\ln^2(\mu^2 \Delta x^2_{\scriptscriptstyle +-}) & \!\!\! = \!\!\! &
4 \pi i \, \theta(\Delta \eta \!-\! \Delta r) \ln[\mu^2 (\Delta \eta^2
\!-\! \Delta r^2)] \; , \qquad \label{log2}
\end{eqnarray}
where $\Delta \eta \equiv \eta_x - \eta_z$ and $\Delta r \equiv \Vert \vec{x}
- \vec{z}\Vert$.

Exploiting relations (\ref{D4prop}-\ref{log2}) allows us to evaluate $I_1$,
\begin{eqnarray}
I_1 = \frac{i 4 H^{D-4}}{(4 \pi)^{\frac{D}2}} \Biggl\{ 
\frac{\Gamma(\frac{D}2 \!-\! 1)}{2 (D\!-\!3) (D\!-\!4)} - \frac13 \ln^2(a_x)
- \frac23 \ln(a_x) + \frac{79}{54} + O\Bigl(\frac1{a_x}\Bigr) \Biggr\} .
\label{I1result}
\end{eqnarray}
For the remaining integrals it is useful to note,
\begin{eqnarray}
\int \!\! d^4z \, {a}_z ^4 \Bigl[ i\Delta_{\scriptscriptstyle ++}(x;z) -
i\Delta_{\scriptscriptstyle +-}(x;z)\Bigr] \!\times\! f(a_z) = -\frac{i}{3 H^2}
\!\! \int_{1}^{a} \!\! \frac{d a_z}{a_z} \Bigl[1 - \frac{{a_z}^3}{a_x^3}\Bigr] 
\!\times\! f(a_z) \; . \label{genform}
\end{eqnarray}
Using (\ref{genform}) with different choices of $f(a_z)$ gives the leading
logarithms results for the other four integrals,
\begin{eqnarray} 
f(a_z) = \frac{H^2 \ln(a_z)}{4 \pi^2} & \Longrightarrow & I_2 \longrightarrow 
-\frac{i}{24 \pi^2} \!\times\! \ln^2(a_x) \; , \label{I2result} \\
f(a_z) = 1 & \Longrightarrow & I_3 \longrightarrow 
-\frac{i}{3 H^2} \!\times\! \ln(a_x) \; , \label{I3result} \\
f(a_z) = -\frac{i \ln(a_z)}{3 H^2} & \Longrightarrow & I_4 \longrightarrow 
-\frac{1}{18 H^4} \!\times\! \ln^2(a_x) \; , \label{I4result} \\
f(a_z) = -\frac{i}{3 H} & \Longrightarrow & I_5 \longrightarrow 
-\frac{1}{9 H^3} \!\times\! \ln(a_x) \; . \label{I5result}
\end{eqnarray}

\end{document}